\documentclass{article}
\usepackage{spconf}
\usepackage{graphicx}
\usepackage{bm}
\usepackage{cite}
\usepackage{amsmath,amssymb,amsfonts,mathrsfs}
\usepackage{bbm}
\usepackage{amsxtra}
\usepackage{threeparttable}
\usepackage{algorithm,algorithmic}
\usepackage{etoolbox,siunitx}
\usepackage{multirow, booktabs}
\usepackage{balance}
\usepackage{textcomp}
\usepackage{xcolor}
\usepackage{url}
\usepackage{pifont}
\usepackage{threeparttable}
\usepackage{hyperref}
\usepackage{makecell}
\usepackage{setspace}
\usepackage{colortbl}
\robustify\bfseries

\usepackage[hang,flushmargin]{footmisc}

\title{Velocity Potential Neural Field\\for Efficient Ambisonics Impulse Response Modeling}
\name{Yoshiki Masuyama,
      Fran\c{c}ois G.\ Germain,
      Gordon Wichern,
      Chiori Hori,
      Jonathan Le Roux}
\address{Mitsubishi Electric Research Laboratories (MERL), Cambridge, MA, USA}

\begin{document}
\ninept
\maketitle
\begin{abstract}
First-order Ambisonics (FOA) is a standard spatial audio format based on spherical harmonic decomposition.
Its zeroth- and first-order components capture the sound pressure and particle velocity, respectively.
Recently, physics-informed neural networks have been applied to the spatial interpolation of FOA signals, regularizing the network outputs based on soft penalty terms derived from physical principles, e.g., the linearized momentum equation.
In this paper, we reformulate the task so that the predicted FOA signal automatically satisfies the linearized momentum equation.
Our network approximates a scalar function called velocity potential, rather than the FOA signal itself.
Then, the FOA signal can be readily recovered through the partial derivatives of the velocity potential with respect to the network inputs (i.e., time and microphone position) according to physics of sound propagation.
By deriving the four channels of FOA from the single-channel velocity potential, the reconstructed signal follows the physical principle at any time and position by construction.
Experimental results on room impulse response reconstruction confirm the effectiveness of the proposed framework.
\end{abstract}


\section{Introduction}
\label{sec:intro}

Sound field reconstruction aims to predict a sound field (typically a sound pressure field) at any point in a given spatial region based on a finite set of measurements~\cite{Fernandez2016,Ueno2025}.
This task has many applications in spatial audio such as immersive audio generation~\cite{Vorlander2015} and sound field control~\cite{Koyama2021}.
Although dense measurements could yield high-quality reconstruction, acquiring them requires an enormous amount of time and effort.
Consequently, various advanced methods have been developed to achieve high-quality reconstruction even from a limited number of measurements~\cite{Meyer2002,Abhayapala2002,verburg2018,laborie2003,Samarasinghe2014,antonello2017,Ueno2018spl,ueno2018iwaenc,lluis2020}.
In this paper, we focus on the spatial interpolation of room impulse responses (RIRs), as they fully characterize the relation between any source and receiver positions and unlock many spatial audio applications~\cite{Koyama2025}.

Traditional RIR interpolation techniques typically leverage physics of sound propagation such as geometrical acoustics~\cite{lee2017,tsunokuni2021} and wave-based acoustics~\cite{antonello2017,Ueno2018spl,ueno2018iwaenc}.
More recently, neural-network-based approaches have shown promising performance thanks to their powerful modeling capability~\cite{bryan2020,ratnarajah2021,pezzoli2022,Della2025diffusionrir}.
In particular, neural fields have been actively used in RIR interpolation, where the network characterizes the sound field as a function of time, source position, and/or microphone position~\cite{luo22naf,richard2022,su22inras}.
Once the neural fields are trained, we can predict RIRs at arbitrary positions in a grid-less manner.
These two directions have been combined with the concept of physics-informed neural networks (PINNs)~\cite{chen2023,pezzoli2023,Karakonstantis2024,sato2024,koyama2024pinnsound}.
A PINN for sound field reconstruction is a neural field whose training is regularized by a physics prior.
Specifically, the derivatives of the predicted sound pressure with respect to the inputs (e.g., time and microphone position) are penalized towards following the governing partial differential equations, e.g., the wave equation~\cite{Karakonstantis2024}.

PINNs have been specialized to model directional RIRs~\cite{merimaa2005}, in particular the (four-channel) first-order Ambisonics (FOA) format~\cite{pidanf}.
FOA~\cite{gerzon1973,daniel1998,arteaga2023} is a widely-used format for spatial audio built on spherical harmonic decomposition~\cite{williams1999}.
Its zeroth-order component corresponds to the sound pressure, and the three remaining first-order components match the particle velocity up to a constant multiplicative factor.
Leveraging these relationships, physics-informed direction-aware neural acoustic field (PI-DANF)~\cite{pidanf} proposes to regularize the neural field training based on physical principles, e.g., the linearized momentum equation~\cite{blackstock2000}.
PI-DANF achieves better reconstruction than a vanilla neural field, but the predicted FOA RIRs are still allowed to deviate from the principles.

\begin{figure}[t!]
\centering
\includegraphics[width=0.99\columnwidth]{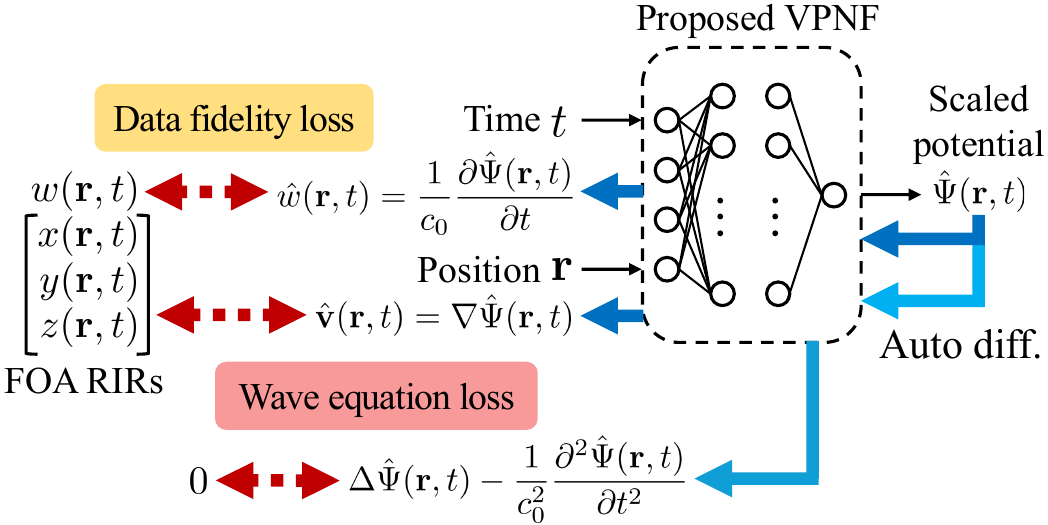}
\caption{Overview of the proposed VPNF, where bold solid arrows indicate automatic differentiation.
We predict the zeroth- and first-order components of Ambisonics by computing the partial derivatives of the potential with respect to time and position, respectively.
}
\label{fig:overview}
\end{figure}

To partially overcome this limitation, we reformulate the task so that the predicted FOA RIRs satisfy the linearized momentum equation by construction.
Under typical assumptions, sound pressure and particle velocity can be computed as partial derivatives of a single scalar function, called velocity potential~\cite{Kinsler2000,Pierce2019}.
The derivatives satisfy the linearized momentum equation by design.
Hence, we aim to model the velocity potential by using a differentiable function.

In this paper, we propose a neural field for velocity potential reconstruction, named a velocity potential neural field (VPNF).
As illustrated in Fig.~\ref{fig:overview}, our neural field outputs a scaled version of the velocity potential.
Then, leveraging automatic differentiation, we recover the four-channel FOA RIRs based on the partial derivatives of the predicted velocity potential with respect to time and microphone position.
This design ensures that the reconstructed sound field follows the linearized momentum equation at any time and position.
In addition, we can incorporate a soft penalty term regularizing the predicted velocity potential towards following the wave equation.
Our experiments confirm the effectiveness of VPNF for FOA RIR interpolation compared with a vanilla neural field and PI-DANF when the number of measurements is limited.

\section{Preliminaries}
\label{sec:preliminaries}
\vspace{-2pt}

\subsection{Problem settings}
\label{sec:settings}

Let the sound pressure and particle velocity be $p(\mathbf{r}, t) \in \mathbb{R}$ and $\mathbf{u}(\mathbf{r}, t) \in \mathbb{R}^3$, respectively, where $\mathbf{r} \in \mathbb{R}^3$ denotes the position in Cartesian coordinates, and $t \in \mathbb{R}$ is the time.
Throughout this paper, we consider a source-free region $\Omega \subset \mathbb{R}^3$ and focus on air sound propagation assuming air is an inviscid fluid.
The sound pressure and particle velocity satisfy the following linearized momentum equation~\cite{blackstock2000}:
\begin{equation}
    \nabla p(\mathbf{r}, t) + \rho_0 \frac{\partial \mathbf{u}(\mathbf{r}, t)}{\partial t} = \mathbf{0}, \label{eq:momentum}
\end{equation}
where $\nabla$ denotes the gradient with respect to $\mathbf{r}$, and $\rho_0$ is the density of the air.
They also follow the continuity equation given by 
\begin{equation}
\rho_0 \nabla \cdot \mathbf{u}(\mathbf{r}, t) + \frac{1}{c_0^2} \frac{\partial p(\mathbf{r}, t)}{\partial t} = 0, \label{eq:continuity}
\end{equation}
where $\nabla \cdot$ is the divergence, and $c_0$ is the sound speed in the air.
The true sound speed and other physical parameters are assumed to be constant and known.

Assuming that air is inviscid, we can define a velocity potential $\Phi(\mathbf{r}, t) \in \mathbb{R}$ in the source-free region $\Omega$ such that~\cite{Kinsler2000,Pierce2019}:
\begin{equation}
    \mathbf{u}(\mathbf{r}, t) = \nabla \Phi(\mathbf{r}, t). \label{eq:velocity}
\end{equation}
By combining this definition with \eqref{eq:momentum}, we obtain the following relation between the velocity potential and sound pressure:
\begin{equation}
    p(\mathbf{r}, t) = -\rho_0 \frac{\partial \Phi(\mathbf{r}, t)}{\partial t}. \label{eq:pressure}
\end{equation}
Then, combining \eqref{eq:continuity}--\eqref{eq:pressure}, we additionally obtain the wave equation for the velocity potential as follows:
\begin{equation}
\Delta \Phi(\mathbf{r}, t) - \frac{1}{c_0^2} \frac{\partial^2 \Phi(\mathbf{r}, t)}{\partial t^2}= 0,
\label{eq:waveeq}
\end{equation}
where $\Delta$ denotes the Laplacian.

Ambisonics is an audio format capturing the spatial characteristics of a sound field based on spherical harmonics~\cite{gerzon1973,daniel1998,arteaga2023}.
Under the SN3D normalization~\cite{daniel2003}, its $W$-channel coincides with the sound pressure measured by an omnidirectional microphone, i.e., $w(\mathbf{r}, t) = p(\mathbf{r}, t)$.
Meanwhile, the $(X, Y, Z)$-channels correspond to the first-order components of spherical harmonics and are proportional to the particle velocity as follows~\cite{merimaa2005}:
\begin{align}
\mathbf{v}(\mathbf{r}, t) &= [x(\mathbf{r}, t); y(\mathbf{r}, t); z(\mathbf{r}, t)], \\
\mathbf{u}(\mathbf{r}, t) &= -\frac{1}{\rho_0 c_0} \mathbf{v}(\mathbf{r}, t),
\end{align}
where $[\cdot ; \cdot]$ denotes vertical concatenation.
Hence, the $(X, Y, Z)$-channels capture the gradient of the velocity potential up to a multiplicative constant.

\subsection{PI-DANF for interpolating FOA RIRs}

RIR interpolation aims to reconstruct spatially-continuous RIRs from sparse  measurements at $\mathbf{r}_d \in \Omega$, where $d = 0, \ldots, D-1$ is the index of the microphone position, with the source location being assumed fixed.
While various signal-processing-based methods have been developed~\cite{antonello2017,lee2017,verburg2018,ueno2018iwaenc,Ueno2018spl}, neural fields have recently gained much attention due to their flexibility and powerful modeling capability~\cite{luo22naf,richard2022,su22inras}.
A neural field for the sound pressure is formulated as follows~\cite{richard2022}:
\begin{equation}
    {p}(\mathbf{r}, t) \approx \hat{p}(\mathbf{r}, t) = \texttt{NF}_\theta(\mathbf{r}, t),
    \label{eq:mononf}
\end{equation}
where $\theta$ are the model parameters.
Once trained, it can predict sound pressure at any time and position in a grid-less manner.

Going beyond RIRs measured by omnidirectional microphones, neural fields have been applied to interpolate FOA RIRs~\cite{danf,pidanf}.
The direction-aware neural field (DANF) predicts the four channels of FOA RIRs instead of only the sound pressure $w(\mathbf{r}, t)$~\cite{danf}:
\begin{equation}
    \hat{w}(\mathbf{r}, t), \hat{\mathbf{v}}(\mathbf{r}, t) = \texttt{DANF}_\theta(\mathbf{r}, t).
\end{equation}
The neural field is optimized to minimize the reconstruction error at the measured positions:
\begin{align}
     \mathcal{L}_\text{data} &= \frac{1}{DL}\sum_{d=0}^{D-1} \sum_{l = 0}^{L-1}  ( |\hat{w}(\mathbf{r}_d, t_l) - w(\mathbf{r}_d, t_l)| \nonumber \\
     &\hspace{90pt} + \|\hat{\mathbf{v}}(\mathbf{r}_d, t_l) - \mathbf{v}(\mathbf{r}_d, t_l)\|_1 ),
     \label{eq:datafidelity-foa}
\end{align}
where $l = 0, \ldots, L-1$ is the sample index of the measurements, and $\| \cdot \|_1$ denotes the $\ell_1$ norm.

Inspired by PINNs~\cite{raissi2019,karniadakis2021}, DANF was extended to incorporate penalty terms derived from the physical principles of sound propagation, as PI-DANF~\cite{pidanf}.
Specifically, the first penalty term is derived from the linearized momentum equation in \eqref{eq:momentum} as follows:
\begin{equation}
\mathcal{L}_\text{momentum} = \mathbb{E}_{\mathbf{r} \in \Omega} \mathbb{E}_{t \in [0, T]} \left\| \nabla \hat{w}(\mathbf{r}, t) - \frac{1}{c_0} \frac{\partial \hat{\mathbf{v}}(\mathbf{r}, t)}{\partial t} \right\|_1,
\label{eq:momentumloss}
\end{equation}
where $T \in \mathbb{R}_+$ is for limiting the time range.
The second term similarly penalizes the discrepancy with the continuity equation in \eqref{eq:continuity}.
These penalty terms incentivize the outputs of PI-DANF to follow the principles of sound propagation even at unmeasured points.
However, it is not guaranteed that the prediction follows the principles strictly.

\vspace{-2pt}
\section{Velocity Potential Neural Field (VPNF)}
\label{sec:proposal}
\vspace{-2pt}

\subsection{Formulation of VPNF}

Our goal is to more explicitly leverage the physical relationship between the sound pressure and particle velocity to train an improved neural field for FOA RIRs.
As shown in Section~\ref{sec:settings}, the $W$-channel corresponds to the sound pressure, and the other channels match the particle velocity.
Then, from \eqref{eq:velocity}--\eqref{eq:pressure}, we obtain the following relation by considering $\Psi(\mathbf{r}, t) = -\rho_0 c_0 {\Phi}(\mathbf{r}, t)$:
\begin{align}
    w(\mathbf{r}, t) &= \frac{1}{c_0} \frac{\partial \Psi(\mathbf{r}, t)}{\partial t}, \label{eq:w-pred} \\
    \mathbf{v}(\mathbf{r}, t) &= \nabla \Psi(\mathbf{r}, t). \label{eq:xyz-pred}
\end{align}
That is, we can predict FOA RIRs $[w(\mathbf{r}, t); \mathbf{v}(\mathbf{r}, t)]$ at any $\mathbf{r} \in \Omega$ and $t \in [0, T]$ by approximating $\Psi(\mathbf{r}, t)$ as a function of $\mathbf{r}$ and $t$ and calculating its derivatives.

We thus propose VPNF that implicitly represents ${\Psi}(\mathbf{r}, t)$ using a neural field, i.e.,
\begin{equation}
     {\Psi}(\mathbf{r}, t) \approx  \hat{\Psi}(\mathbf{r}, t) = \texttt{VPNF}_\theta(\mathbf{r}, t).
\end{equation}
Since it is impractical to measure the ground-truth velocity potential, we train VPNF using the following data-fidelity term, matching its derivatives to the FOA RIRs at the measured positions:
\begin{align}
     \mathcal{L}_\text{data} &= \frac{1}{DL}\sum_{d=0}^{D-1} \sum_{l = 0}^{L-1}  \Bigl( \Bigl|\frac{1}{c_0}\frac{\partial \hat{\Psi}(\mathbf{r}, t)}{\partial t} - w(\mathbf{r}_d, t_l) \Bigr| \Bigr. \nonumber \\
     &\hspace{80pt} + \Bigl. \Bigr\| \nabla \hat{\Psi}(\mathbf{r}_d, t_l) - \mathbf{v}(\mathbf{r}_d, t_l)\Bigr\|_1 \Bigr),
     \label{eq:datafidelity-vpnf}
\end{align}
where the partial derivatives are computed using automatic differentiation in deep learning frameworks.
In addition, we can incorporate another penalty term derived from the wave equation in \eqref{eq:waveeq}: 
\begin{equation}
     \mathcal{L}_\text{wave} =  \mathbb{E}_{\mathbf{r} \in \Omega} \mathbb{E}_{t \in [0, T]} \Bigl| \Delta \hat{\Psi}(\mathbf{r}, t) - \frac{1}{c_0^2} \frac{\partial^2 \hat{\Psi}(\mathbf{r}, t)}{\partial t^2}  \Bigr|.
     \label{eq:waveeqloss}
\end{equation}
We note that this penalty term can be computed at arbitrarily sampled positions $\mathbf{r} \in \Omega$ and $t \in [0, T]$ in a grid-less manner.

The FOA RIRs predicted by VPNF satisfy the linearized momentum equation in \eqref{eq:momentum} at any time and position thanks to their derivation in \eqref{eq:w-pred}--\eqref{eq:xyz-pred}.
Meanwhile, PI-DANF separately predicts the $W$- and $(X, Y, Z)$-channels and exploits that equation only as a penalty term~\cite{pidanf}.
It is thus not guaranteed that the FOA RIRs predicted using PI-DANF satisfy the linearized momentum equation exactly.
We would argue that VPNF more strictly integrates the neural field with the physical principles of sound propagation.
Conversely, the penalty term in \eqref{eq:waveeqloss}, derived from the continuity equation in \eqref{eq:continuity}, means that VPNF, like PI-DANF, still does not guarantee that the predicted FOA RIRs exactly follow the continuity equation in contrast to the linearized momentum equation.
It would require a more complex design to ensure the prediction simultaneously follows both equations.
We leave such a development to future work.

\subsection{Network architecture and training setup}

To realize VPNF, we mainly follow the network architecture and training strategy of PI-DANF~\cite{pidanf}.
The network comprises a modified multi-layer perceptron (MLP) with SIREN activations~\cite{sitzmann2020}, where the periodic activation provides well-behaved derivatives~\cite{Karakonstantis2024}.
Then, VPNF is realized as follows:
\begin{equation}
    \texttt{VPNF}_\theta(\mathbf{r}, t)
    = \texttt{ModifiedMLP}^{(1)}_\theta(\mathbf{r}, c_0 t),
\end{equation}
where the superscript denotes the output dimension of the modified MLP, and the final layer outputs a scalar without any activation.
Here, we convert the time $t$ to distance $c_0 t$ to align the scale of the network inputs.
Furthermore, we experimentally find a simple modification could improve the performance as follows:
\begin{equation}
    \texttt{VPNF}_{\theta}^+(\mathbf{r}, t)
    = [c_0 t; \mathbf{r}]^\mathsf{T} \, \texttt{ModifiedMLP}^{(4)}_\theta(\mathbf{r}, c_0 t),
    \label{eq:vpnfplus}
\end{equation}
where the output dimension of the modified MLP is four.
We expect this parametrization makes the gradient of $\Psi(\mathbf{r}, t)$ close to $\texttt{ModifiedMLP}^{(4)}_\theta(\mathbf{r}, c_0 t)$ except for the multiplicative factor $c_0$, especially around $(\mathbf{r}, t) = (\mathbf{0}, 0)$.
It may be beneficial as the modified MLP has worked well for modeling single-channel~\cite{Karakonstantis2024} and FOA RIRs~\cite{pidanf}.

The network is optimized to minimize the data fidelity term in \eqref{eq:datafidelity-vpnf} or its sum with the penalty term in \eqref{eq:waveeqloss}.
When incorporating the penalty term, we balance the two terms adaptively~\cite{xiang2022,Karakonstantis2024}:
\begin{equation}
    \!\! \mathcal{L}_\text{all} = \frac{1}{2 \epsilon_\text{data}^2} \mathcal{L}_\text{data} + \frac{1}{2 \epsilon_\text{wave}^2} \mathcal{L}_\text{wave} + \log (\epsilon_\text{data} \epsilon_\text{wave}),
    \label{eq:adaptive}
\end{equation}
where $\epsilon_\text{data}$ and $\epsilon_\text{wave}$ are the weights for each term, and these parameters are also updated by backpropagation.
The position $\mathbf{r} \in \Omega$ and time $t \in [0, T]$ for \eqref{eq:waveeqloss} are randomly sampled at each iteration.

\section{Experiments}

\subsection{Dataset and experimental setup}

\begin{figure}[t!]
\centering
\includegraphics[width=0.99\columnwidth]{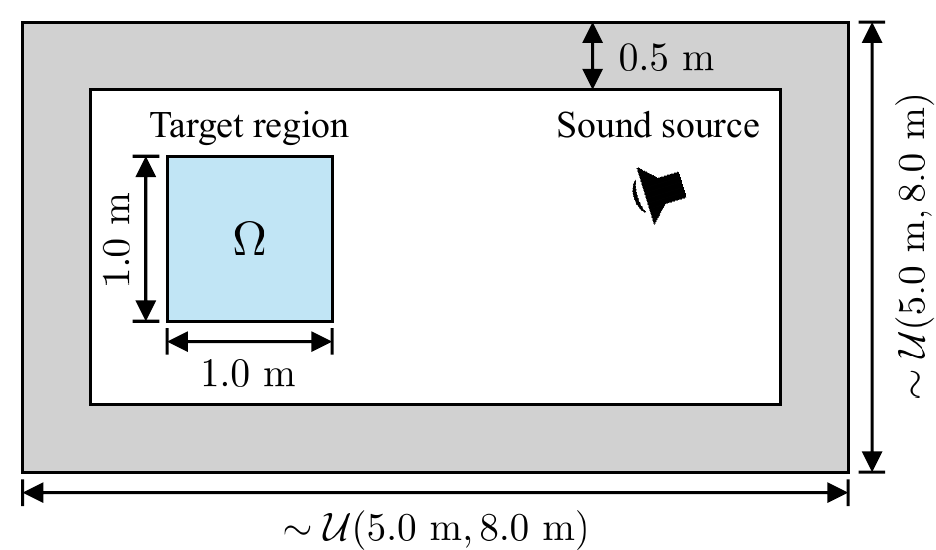}
\caption{Overhead view of the room geometry for simulation.}
\label{fig:exproom}
\end{figure}

The proposed VPNF is evaluated on FOA RIRs simulated with \texttt{HARP}%
\footnote{\url{https://github.com/whojavumusic/HARP}}
~\cite{saini2024}, an extension of \texttt{Pyroomacoustics}~\cite{Scheibler2018PyRoom} for Ambisonics.
The direct sound and early reflections up to $100$~ms are simulated at $8$~kHz in ten shoebox-shaped rooms with randomly sampled room dimensions and surface materials.
An example room configuration is illustrated in Fig.~\ref{fig:exproom}, where the room length/width and height were randomly sampled from $[5.0, 8.0)$ m and $[2.5, 4.5)$ m, respectively.
The (blue) target region $\Omega$ is a cube with $1.0$ m size.
The cube and a sound source are randomly placed within the room excluding a $0.5$ m buffer near the walls (see shaded area in Fig.~\ref{fig:exproom}).
RIRs were simulated on a grid of $5$ cm width in Cartesian coordinates, resulting in \num{9261} measurements.

We assess the performance under two different measurement setups.
The first one is the reconstruction from a set of FOA RIRs randomly sampled from the \num{9261} measurements.
We randomly choose a fixed set of $\{30, 50, 70, 100, 150, 200\}$ measurements to train the neural fields and use $50$ distinct measurements for validation.
The evaluation is performed on the remaining \num{9011} positions using the model with the best 
normalized mean squared error (NMSE) for the $W$-channel on the validation set.
In the second setup, we train the neural field on a fixed set of $\{100, 200\}$ FOA RIRs measured at the surface of the cube.
This setup is more challenging than the first one, as we cannot access measurements within the cube.
We again use $50$ additional data measured on the surface for validation and evaluate the performance of the checkpoint with the best validation score on the remaining \num{9011} positions.

Our network and training configuration follow those used in PI-DANF~\cite{pidanf}, which is based on a prior PINN for omnidirectional RIRs%
\footnote{\url{https://github.com/xefonon/RIRPINN/tree/main}}~\cite{Karakonstantis2024}.
VPNF and its variants consist of 3 hidden layers with $512$ hidden units.
They are optimized with the Adam optimizer and cosine annealing.
During training, we randomly sample $250$ discrete times $t_l$ from all the training-data positions to compute the data fidelity term in \eqref{eq:datafidelity-vpnf}.
When incorporating the wave-equation-based penalty term in \eqref{eq:waveeqloss}, we randomly sample $25,000$ pairs of $\mathbf{r}$ and $t$ by Latin hypercube sampling~\cite{lhs}.
We set the initial values of $\epsilon_\text{data}$ and $\epsilon_\text{wave}$ to $1.0$ and $0.1$, respectively.
The networks are trained for \num{100000} iterations.

The performance is evaluated by NMSE in dB scale for the $W$-channel and $(X, Y, Z)$-channels:
\begin{equation}
    \text{NMSE}(s, \hat{s}) = 10 \log_{10} \left(
    \frac{\sum_{\tilde{d}=0}^{\tilde{D}-1} \|\hat{\mathbf{s}}(\mathbf{r}_{\tilde{d}}) - \mathbf{s}(\mathbf{r}_{\tilde{d}})\|_2^2}{\sum_{\tilde{d}=0}^{\tilde{D}-1} \|\mathbf{s}(\mathbf{r}_{\tilde{d}})\|_2^2}\right),
\end{equation}
where $s$ is any of the four channels of FOA RIRs, $\tilde{d} = 0, \ldots, \tilde{D}-1$ indexes evaluation positions, $\mathbf{s}(\mathbf{r}_{\tilde{d}}) = [s(\mathbf{r}_{\tilde{d}}, t_0), \ldots, s(\mathbf{r}_{\tilde{d}}, t_{L-1})]^\mathsf{T}$, and $\|\cdot\|_2$ denotes the $\ell_2$ norm.
Another metric is the Pearson’s correlation coefficient between the reference and predicted RIRs at each channel, motivated by its correlation with perceptual localization accuracy~\cite{ren2024icassp}:
\begin{equation}
    \!\!\text{PCC}(s, \hat{s})
    \!=\!
    \frac{1}{\tilde{D}}
    \sum_{\tilde{d}=0}^{\tilde{D}-1} 
     \frac{[\hat{\mathbf{s}}(\mathbf{r}_{\tilde{d}}) - \hat{\underline{\mathbf{s}}}(\mathbf{r}_{\tilde{d}})]^\mathsf{T}
     [{\mathbf{s}}(\mathbf{r}_{\tilde{d}}) - \underline{\mathbf{s}}(\mathbf{r}_{\tilde{d}})]}{
      \|\hat{\mathbf{s}}(\mathbf{r}_{\tilde{d}}) - \hat{\underline{\mathbf{s}}}(\mathbf{r}_{\tilde{d}})\|_2
     \|{\mathbf{s}}(\mathbf{r}_{\tilde{d}}) - \underline{\mathbf{s}}(\mathbf{r}_{\tilde{d}})\|_2
     },\!\!
\end{equation}
where $\hat{\underline{\mathbf{s}}}(\mathbf{r}_{\tilde{d}})$ and $\underline{\mathbf{s}}(\mathbf{r}_{\tilde{d}})$ are the time average of $\hat{s}(\mathbf{r}_{\tilde{d}}, t_l)$ and $s(\mathbf{r}_{\tilde{d}}, t_l)$, respectively.
We took the average of these channel-wise metrics over the $(X, Y, Z)$-channels.

\begin{figure}[t!]
\centering
\includegraphics[width=0.99\columnwidth]{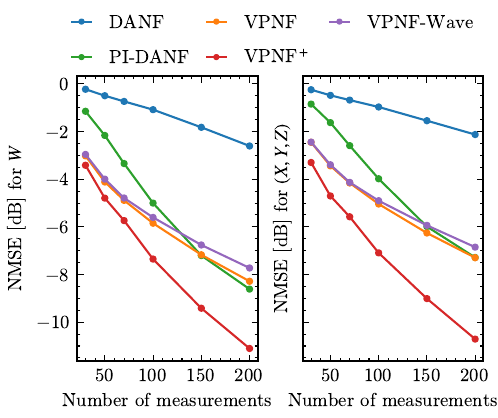}
\caption{NMSE [dB] averaged over ten rooms with different numbers of measurements.
}
\label{fig:nmse}
\end{figure}

\begin{figure}[t!]
\centering
\includegraphics[width=0.99\columnwidth]{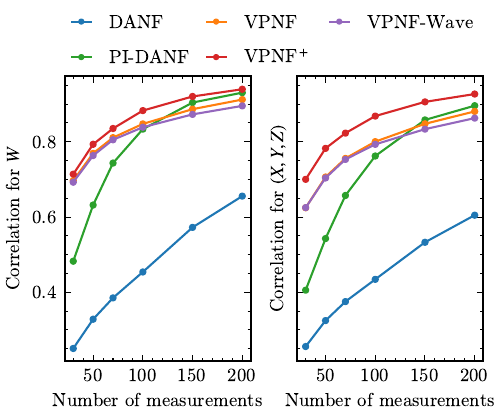}
\caption{Average Pearson’s correlation coefficients for the $W$- and $(X, Y, Z)$-channels.}
\label{fig:pcc}
\end{figure}

\subsection{Results}

Figure~\ref{fig:nmse} shows the NMSE for the $W$- and $(X, Y, Z)$-channels with different numbers of measurements under the first condition, where the measurements were randomly sampled from the target region.
Here, DANF~\cite{danf} denotes a neural field for FOA RIRs trained only with the data-fidelity term in \eqref{eq:datafidelity-foa}.
PI-DANF additionally exploits the physical principles in \eqref{eq:momentum}--\eqref{eq:continuity} as soft penalties~\cite{pidanf}. 
Consistently with prior trends in \cite{pidanf}, PI-DANF reliably outperforms DANF as a result of regularizing the network outputs according to the underlying physics.
The proposed VPNF achieves lower NMSE than all baselines when the number of measurements is less than or equal to $100$.
Its performance is however comparable with PI-DANF when we have more measurements.
These results demonstrate the effectiveness of enforcing the linearized momentum equation by modeling velocity potential, especially under the more challenging conditions, i.e., with fewer of measurements.
Among variants of VPNF, we find that VPNF$^+$ (see Eq.~\eqref{eq:vpnfplus}) substantially improves the performance.
Interestingly, using Eq.~\eqref{eq:waveeqloss} as a soft penalty, i.e., VPNF-Wave, shows little improvement in our tests.
We will analyze this point further in future work.
Similar tendencies are found in the Pearson’s correlation coefficient results, as seen in Fig.~\ref{fig:pcc}.

Table~\ref{tab:surface} summarizes the reconstruction performance from measurements on the surface of the cube.
Under this more challenging condition, VPNF and its variants show a clear edge over all baselines for all metrics.

\begin{table}[t]
\sisetup{
detect-weight,
mode=text,
tight-spacing=true,
round-mode=places,
round-precision=2,
table-format=1.2,
table-number-alignment=center
}
  \vskip -2mm
  \caption{
    Reconstruction results from $\{100, 200\}$ measurements on the surface of the target region.
  }
  \vskip 1mm
  \label{tab:surface}
  \centering
  \resizebox{\linewidth}{!}{
  \setlength{\tabcolsep}{4pt}
  \begin{tabular}{l*{4}{S[table-format=2.2]}*{4}{S}}
    \toprule
     & \multicolumn{4}{c}{NMSE [dB] ($\downarrow$)} & \multicolumn{4}{c}{Correlation ($\uparrow$)} \\
     \cmidrule(lr){2-5} \cmidrule(lr){6-9} 
     & \multicolumn{2}{c}{$W$-ch} & \multicolumn{2}{c}{$(X, Y, Z)$-ch} & \multicolumn{2}{c}{$W$-ch} & \multicolumn{2}{c}{$(X, Y, Z)$-ch} \\
     \cmidrule(lr){2-3} \cmidrule(lr){4-5} \cmidrule(lr){6-7} \cmidrule(lr){8-9} 
      & {100} & {200} & {100} & {200} & {100} & {200} & {100} & {200}  \\
     \midrule 
     DANF~\cite{danf} & -0.38 & -0.37 & -0.40 & -0.43 & 0.31 & 0.38 & 0.31 & 0.36 \\
     PI-DANF~\cite{pidanf} & -2.03 & -3.87 & -1.52 & -2.99 & 0.62 & 0.77 & 0.53 & 0.69 \\
     VPNF & -3.97 & -5.04 & -3.26 & -4.15 & 0.75 & 0.81 & 0.69 & 0.75 \\
     VPNF-Wave & -3.83 & -4.74 & -3.29 & -4.09 & 0.74 & 0.79 & 0.69 & 0.74\\
     VPNF$^+$ & \bfseries-4.50 & \bfseries-6.31 & \bfseries-4.24 & \bfseries-5.92 & \bfseries0.78 & \bfseries0.86 & \bfseries0.76 & \bfseries0.83 \\
    \bottomrule
  \end{tabular}
  }
  \vskip -2mm
\end{table}

\vskip -2mm
\section{Conclusion}

We presented VPNF, a PINN that models the velocity potential field and predicts FOA RIRs by taking the output gradient with respect to the network inputs (i.e., time and microphone position).
By design, the predicted FOA RIRs are guaranteed to satisfy the linearized momentum equation at any time and position.
We also tested the additional physics-informed penalty term derived from the wave equation for the velocity potential.
Our experiments demonstrate the effectiveness of VPNF, especially when the number of measurements is limited.
Future work will explore few-shot adaptation of a pre-trained VPNF to new rooms to minimize the required measurements.

\clearpage
\balance

\let\oldthebibliography\thebibliography
\renewcommand{\thebibliography}[1]{%
  \oldthebibliography{#1}%
  \footnotesize
  \setlength{\itemsep}{0.1pt}%
  \setlength{\parskip}{0.1pt}%
}
\bibliographystyle{IEEEtran}
\bibliography{refs}
\end{document}